\documentclass[aps,prd,twocolumn,showpacs,nofootinbib,amsmath,amssymb,amsfonts]{revtex4}
\usepackage{graphicx,epsfig,rotate,psfig}   
\usepackage{dcolumn}      
\usepackage{bm}           
\usepackage{amsmath}
\usepackage{amssymb}
\usepackage{latexsym}

\def\setC{\mathbb{C}}

\def\setR{\mathbb{R}}

\newcommand{\cav}{\mathrm{cav}}

\newcommand{\ppn}{\mathrm{PPN}}

\newcommand{\mat}{\mathrm{m}}
\newcommand{\de}{\mathrm{DE}}

\newcommand{\dd}{\mathrm{d}}

\newcommand{\ie}{\textsl{i.e.~}}

\newcommand{\etal}{\textsl{et al.~}}

\newcommand{\GReCO}{${\cal G}\setR\varepsilon\setC{\cal O}$}

\def\spose#1{\hbox to 0pt{#1\hss}}
\def\lta{\mathrel{\spose{\lower 3pt\hbox{$\mathchar"218$}}
     \raise 2.0pt\hbox{$\mathchar"13C$}}}
\def\gta{\mathrel{\spose{\lower 3pt\hbox{$\mathchar"218$}}
     \raise 2.0pt\hbox{$\mathchar"13E$}}}

\begin{document}

\title{Testing for $w<-1$ in the Solar System}

\author{J\'er\^ome Martin}
 \email{jmartin@iap.fr}
 \affiliation{Institut d'Astrophysique de Paris,
              \GReCO-CNRS, UMR 7095, Universit\'e Pierre et Marie
 Curie, 98bis boulevard Arago, 75014 Paris, France}

\author{Carlo Schimd}
 \email{schimd@iap.fr}
 \affiliation{Institut d'Astrophysique de Paris,
              \GReCO-CNRS, UMR 7095, Universit\'e Pierre et Marie
 Curie, 98bis boulevard Arago, 75014 Paris, France}

\author{Jean-Philippe Uzan}
 \email{uzan@iap.fr}
 \affiliation{Institut d'Astrophysique de Paris,
              \GReCO-CNRS, UMR 7095, Universit\'e Pierre et Marie
 Curie, 98bis boulevard Arago, 75014 Paris, France}

\date{\today}


\begin{abstract}
In scalar-tensor theories of gravity, the equation of state of dark
energy, $w$, can become smaller than $-1$ without violating any energy
condition. The value of $w$ today is tied to the level of deviations
from general relativity which, in turn, is constrained by solar system
and pulsars timing experiments. The conditions on these local
constraints for $w$ to be significantly less than $-1$ are
established. It is demonstrated that this requires to consider
theories that differ from the Jordan-Fierz-Brans-Dicke theory and that
involve either a steep coupling function or a steep potential. It is
also shown how a robust measurement of $w$ could probe scalar-tensor
theories.
\end{abstract}

\maketitle


Various observations indicate that the expansion of our Universe is
presently accelerated~\cite{SNIa}. While still debated, this
conclusion appears to be more and more robust and, as a consequence,
the discussion has now mainly shifted to explaining the cause of this
acceleration. The property of the effective equation of state inferred
from the observations, $w$, is a key issue in this investigation. In
particular, showing that $w\neq -1$ and/or ${\rm d}w/{\rm d}z\neq 0$
($z$ being the redshift) would exclude a cosmological constant,
probably the most natural candidate and, hence, would have drastic
implications for fundamental physics. Recently, various observations
have pointed towards the conclusion that
$w<-1$~\cite{dataneg}. Although far from being settled, this would
also have important consequences since such an equation of state
cannot be achieved by quintessence models~\cite{quint} for which
$-1\leq w\leq 1$. ``Phantom models''~\cite{fantom}, consisting in a
scalar field with a minus sign in front of the kinetic term, are very
often advocated in order to explain $w<-1$. These models are plagued
by various theoretical problems such as, for instance, their stability
when interactions with other fields are taken into account.

\par

However, another route can be investigated since theories where the
gravity sector is modified, \ie where gravity is no longer described
by general relativity (GR), can also entertain $w<-1$ even if the
matter sector is described in a standard fashion. The prototype of
such a theory is a scalar-tensor theory of gravity which is both
well-defined and well-motivated~\cite{def} as they arise as the low
energy limit of string theory. It is worth recalling that they are
equivalent to theories where the gravitational action is given by an
arbitrary function of the Ricci scalar and also encompass the
Jordan-Fierz-Brans-Dicke (JFBD) theory as a particular case.  As
already mentioned, having $1+w$ negative is linked to the fact that
the gravity sector is modified but such modifications are strongly
constrained by solar system and pulsars timing experiments. Therefore,
one can hope to use these local tests to track the real nature of the
dark sector. In this letter, we investigate these issues in general
scalar-tensor theories.


In the Jordan frame, \ie in the (physical) frame where the
experimental data have their usual interpretation, scalar-tensor
theories are described by the action~\cite{gefpol}
\begin{eqnarray}
\label{action}
S &=& \frac{1}{16\pi G_*}\int {\rm d}^4x\sqrt{-g}\bigl[F\left(\varphi
  \right)R-Z\left(\varphi \right)g^{\mu \nu }\partial _{\mu }\varphi
  \partial _{\nu }\varphi \nonumber \\ & & -2U\left(\varphi
  \right)\bigr]+S_\mat\left[\Psi _\mat,g_{\mu \nu }\right]\, ,
\end{eqnarray}
which depends on three arbitrary functions $F\left(\varphi \right)$,
$Z\left(\varphi \right)$ and the potential $U\left(\varphi \right)$,
only two of which are independent. $G_*$ is a constant, different from
the gravitational constant measured in a Cavendish experiment,
$G_\cav$, as will be discussed in more details below. In the
following, the matter action, $S_{_\mat}[\Psi _{_\mat}, g_{\mu \nu}]$,
describes a pressure-less perfect fluid.

\par

In the particular case of a Friedmann-Lema\^{i}tre-Roberston-Walker
(FLRW) Universe, choosing $Z\left(\varphi \right)=1$, the field
equations reduce to
\begin{eqnarray}
& & 3\left(H^2+\frac{{\cal K}}{a^2}\right)=8\pi G_* \frac{\rho_{_\mat}
 }{F}+\frac{\dot{\varphi }^2}{2F}-3H\frac{\dot{F}}{F}+\frac{U}{F}\, ,
\label{st1} \\
& & -\left(2\frac{\ddot{a}}{a}+H^2+\frac{{\cal K}}{a^2}\right) =
\frac{\dot{\varphi
  }^2}{2F}+\frac{\ddot{F}}{F} +2H\frac{\dot{F}}{F}-\frac{U}{F}\, ,
\label{st2}
\end{eqnarray}
with $H\equiv \dot{a}/a$, $a$ being the scale factor and ${\cal K}$
the curvature of the spatial sections. The energy density of matter
scales as $a^{-3}$. These equations should be compared to $3(H^2+{\cal
K}/a^2)=8\pi G_\cav (\rho _{_\mat} +\rho _{_\de})$ and
$-\left(2\ddot{a}/a+H^2+{\cal K}/a^2\right)=8\pi G_\cav p_{_\de}$,
since the equation of state of dark energy is experimentally inferred
from the expansion history of the Universe by using the standard
Friedmann equations of GR. It follows that
$3\Omega_{_\de}w\equiv-1+\Omega_{\mathcal{K}}+2q$ where $q\equiv
-a\ddot{a}/(\dot{a})^2$ is the acceleration parameter,
$\Omega_\mathcal{K}\equiv -\mathcal{K}/a^2H^2$ and that the dark
energy density parameter is defined as $\Omega_{_\de}(z)\equiv
H^2/H_0^2 - \Omega_{\mat}^0(1+z)^3-\Omega_{\mathcal{K}}^0(1+z)^2$,
where here and in the rest of this letter the subscript ``$0$''
denotes the present day value of the corresponding quantity. Clearly,
identifying $\rho _{_\de} $ and $p_{_\de}$ from Eqs.~(\ref{st1}) and
(\ref{st2}) leads to the conclusion that, in scalar-tensor theories,
$w$ needs not to be positive definite, depending on the choice of
$F$. This was already stressed by various works~\cite{precur} and
worked out in the particular case of JFBD theories in
Refs.~\cite{jfbd}.

\par

The present values of $F$ and its derivatives are constrained by the
solar system and pulsars experiments. In order to reveal the link
between $w$ and these constraints, it is more convenient to work in
the Einstein frame. The Einstein frame metric, $g_{\mu \nu}^*$ (all
Einstein frame quantities will be denoted with a star), is related to
the physical metric through the conformal transformation $g_{\mu\nu}^*
= F\left(\varphi \right)g_{\mu\nu}$.  Setting
$F(\varphi)=A^{-2}(\varphi_*)$, the scale factors and cosmic times in
both frames are related by $a=Aa_*$ and $\dd t =A\dd t_*$, so that the
Hubble parameters are linked by $AH = H_*\left(1 + \alpha
\varphi'_*\right)$ with $H_*=\dd\ln a_*/\dd t_*$ and where a prime
denotes a derivative with respect to the Einstein frame number of
$e$-folds, $N_*\equiv \ln(a_*/a_{*0})$.  Deviations from GR are
usually described by two parameters, $\alpha $ and $\beta $, which are
the first and second field derivative of the coupling function
$A(\varphi_*)$, namely
\begin{equation}
 \alpha \equiv \frac{\dd\ln A}{\dd\varphi_*},\quad \beta \equiv
 \frac{\dd\alpha}{\dd\varphi_*}\, .
\end{equation}
In this framework, GR is characterized by $\alpha=\beta=0$ while the
JFBD theory corresponds to a constant $\alpha^2\equiv(2\omega_{_{\rm
BD}}+3)^{-1}$ and $\beta =0$. As mentioned before, in scalar-tensor
theories, the Newton constant obtained in a Cavendish experiment
differs from $G_*$ and is given by
\begin{equation}
\label{defG}
 G_\cav = G_*A_0^2(1+\alpha_0^2)\, .
\end{equation}
The parameters $\alpha_0$ and $\beta_0$ are constrained by various
experiments. If we define $\bar\gamma \equiv\gamma^{_\ppn} - 1$ and
$\bar\beta\equiv \beta^{_\ppn} - 1$, where $\gamma^{_\ppn}$ and
$\beta^{_\ppn}$ are the post-Newtonian parameters~\cite{will}, related
to $\alpha_0$ and $\beta_0$ by $\bar\gamma =
-{2\alpha_0^2}/{(1+\alpha^2_0)}$ and $2\bar\beta =
{\beta_0\alpha_0^2}/{(1+\alpha_0^2)^2}$, then the perihelion shift of
Mercury implies $\left \vert 2\bar\gamma - \bar\beta\right \vert
<3\times10^{-3}$ while the Lunar Laser Ranging experiment sets
$4\bar\beta - \bar\gamma = -(0.7\pm1)\times10^{-3}$. On the other
hand, a bound on $\bar\gamma$ alone is set from the time delay
variation to the Cassini spacecraft near solar conjunction, namely
$\bar\gamma = (2.1\pm2.3)\times10^{-5}$, see Ref.~\cite{pulsar} for a
review. We conclude that
\begin{equation}
\alpha_0^2< 10^{-5}, \quad \beta_0\gtrsim-4.5\, ,
\end{equation}
where the lower bound on $\beta_0$ arises from pulsar timing
experiments~\cite{pulsar}. Note, however, that we cannot consider
arbitrarily large values of $\beta_0$ since then the post-Newtonian
approximation scheme would breakdown. In this case, the upper
constraints should be reanalyzed. We thus loosely assume that
$\beta_0\lesssim100$.

\par

Working out the Friedmann equations in Einstein frame~\cite{gefpol},
we obtain that the equation of state is given by
\begin{widetext}
\begin{eqnarray}\label{eqw}
 3\Omega_{_\de}w= -1 + \Omega_{\cal K}
   +(1+\alpha\varphi_*')\left[3\frac{A^2}{A^2_0(1+\alpha_0^2)}-2\right]\Omega_\mat
   - 2(1+\alpha\varphi_*')\Omega_{_\de} +
   2\frac{\alpha\varphi_*'(2+\alpha\varphi_*')+\varphi_*^{\prime2}}{1+\alpha\varphi_*'}
   -
   2\frac{\alpha\varphi_*''+\beta\varphi_*^{\prime2}}{(1+\alpha\varphi_*')^2},
\end{eqnarray}
\end{widetext}
where $\Omega_\mat \equiv 8\pi G_\cav \rho_\mat/3H^2$ so that
$\Omega_{_\de}+\Omega_\mat+\Omega_{\cal K}=1$. In the limit of a
minimally coupled scalar field, it reduces to the standard relation
$1+w = 2\varphi_{*}^{\prime2}/3\Omega_{_\de}$.

\par

At this point, it is of utmost importance to stress that the value of
$\varphi_*^{\prime}$ is not free. Indeed, in the Einstein frame, the
Friedmann equation reads~\cite{gefpol}
\begin{eqnarray}
 H_*^2 \left(3-\varphi_*^{\prime 2}\right)=-3\frac{\cal K}{a_*^2}+8\pi
   G_*\rho_{\mat*} + 2V(\varphi_*)\label{einsteinEF1}\, ,
\end{eqnarray}
where $V(\varphi_*)=U/(2F^2)$, and the positivity of the energy
density of matter implies that $|\varphi_*^{\prime}|<\sqrt{3}$, as
long as $\Omega_{\cal K}\ll\Omega_\mat$ (in the following, we
assume that $\Omega_{\cal K}=0$).

\par

The expression~(\ref{eqw}) for $w$ is completely general but very
intricate and hence not so illuminating. However, taking into account
that $\alpha_0$ has to be small and that $\varphi_*'$ has to be
bounded by $\sqrt{3}$ and, therefore, that $\alpha
_0\varphi_{*0}^{\prime}$ has to be small as well, the present day
value of the equation of state simplifies considerably and reduces to
\begin{eqnarray}\label{wapp}
 3\Omega_{_{\de}}^0(1+w_0) \simeq 2(1-\beta_0)\varphi_{*0}^{\prime2}
   -2\alpha_0\varphi_{*0}''.
\end{eqnarray}
This formula turns out to be our main result. The contribution of
$\beta_0$ arises from the term $\ddot F/F$ in the right hand side of
Eq.~(\ref{st2}). In fact, Eq.~(\ref{wapp}) shows that $w_0<-1$ is
always possible provided $\beta_0>1$ (in this case, even if the
slow-roll approximation is satisfied, \ie $\varphi''_{*0}\ll
\varphi'_{*0}$) and/or $\alpha\varphi''_{*0}$ positive and large
compared to $\varphi_{*0}^{\prime2}$. Both regimes cannot be reached
in the case of JFBD theories (except if the time variation of $G_\cav$
is large, see below) and exist even in the limit
$\alpha_0\rightarrow0$ so that all local tests can be satisfied.

The amplitude of $w_0$ depends on the value of $\varphi'_{*0}$ and
$\varphi''_{*0}$. Independently of any dynamics, these two quantities
are constrained by the bounds on the time variation of the
gravitational constant~\cite{ctes}. Defining
\begin{equation}
 \frac{\dd \ln G_\cav}{\dd t} \equiv \sigma H\,,
 \quad
 \frac{\dd^2 \ln G_\cav}{\dd t^2} \equiv \xi H^2\,,
\end{equation}
the parameter $\sigma _0$ is bounded by $|\sigma_0|
<5.86\times10^{-2}h^{-1}$. There is no stringent bounds on $\xi_0$ but
we can estimate that, since $\sigma_0$ has been ``measured'' during a
period of about 20 years, we have $|\xi_0 H_0^2|\lesssim |\dot
G/G|_0/(20$~yr). This implies that $|\xi_0|\lesssim 5\times
10^{8}h^{-1}\sigma_0\sim 2.5\times10^7h^{-2}$. Using Eq.~(\ref{defG}),
one can then express $\varphi'_{*}$ and $\varphi''_{*}$ in terms of
the parameters $\sigma $ and $\xi$. As long as
$\beta\not=-(1+\alpha^2)$, a case we shall discuss later, one arrives
at
\begin{eqnarray}
\label{phiprim0}
\varphi_*'
 =\frac{\sigma}{2\alpha}\left(1+\frac{\beta}{1+\alpha^2}-\frac{\sigma
 }{2}\right)^{-1} \, ,
\end{eqnarray}
while the second derivative reads
\begin{widetext}
\begin{eqnarray}
\varphi_{*}'' &=& \frac{(1+\alpha\varphi_*')^2}{2\alpha
    \left[1+\beta/\left(1+\alpha ^2\right)\right]}\xi
    -\varphi_*'\biggl\lbrace
    \varphi_*'\left[\frac{\beta}{\alpha}-\alpha
    +\frac{1}{1+\alpha^2+\beta } \left(\frac{\dd\beta}{\dd\varphi_*} -
    \frac{2\alpha \beta}{1+\alpha^2}\right) \right] \nonumber\\ &&
    +(1+\alpha\varphi_*')^2\left[\Omega_{_\de}
    -\left(3\frac{A^2/A_0^2}{1+\alpha_0^2}-2
    \right)\frac{\Omega_\mat}{2}-1 \right]-\varphi_*^{\prime2}
    \biggr\rbrace.
\end{eqnarray}
\end{widetext}
Interestingly enough, we see that $\varphi _{*0}'' $, and hence $w_0$,
depends on $\xi _0$ and on the derivative of the $\beta $
function. The latter is not constrained and we will assume that
$\dd\beta/\dd\varphi_*\lesssim\mathcal{O}(100)$ for the same reason as
for $\beta$. The above formula is rather complicated but, given the
previous constraints, we simply have
\begin{eqnarray}
\label{phippapprox}
\alpha \varphi_*'' &\simeq & \frac{\xi }{2+2\beta/\left(1+\alpha
^2\right)} -\frac{\beta }{1+\beta/\left(1+\alpha
^2\right)}
\nonumber \\
& & \times
\left(1+\frac{\beta }{1+\alpha ^ 2}\frac{1-\alpha
^2}{1+\alpha ^2}\right)\varphi _*^{\prime 2}\, .
\end{eqnarray}
In particular, the unknown term $\dd\beta/\dd\varphi_*$ does not
appear in this approximation.

\par

As mentioned above, the previous considerations are independent from
the dynamics. However, from a model building point of view,
$\varphi_*''$ and $\varphi_*'$ are not independent once the potential
$V(\varphi_*)$ is chosen. They are related through the Klein-Gordon
equation which reads
\begin{equation}
 \frac{2(X+1)}{3-\varphi_*^{\prime2}}\,\varphi_{*}''
 +{(2+X)}\varphi'_*
 =-({\alpha X +\alpha_{\rm v}})\, ,
\end{equation}
with $X\equiv\Omega_\mat/\Omega_{_{\rm U}}$, where $\rho _{_{\rm
U}}\equiv 2U/(16\pi G_*)$ and $\alpha_{\rm v}\equiv \dd\ln
V/\dd\varphi_*$. When the kinetic energy of the scalar field is
negligible, we have $\Omega_{_{\rm U}}\sim \Omega _{_{\rm DE}}$.

\par

\begin{figure}
\centerline{\epsfig{figure=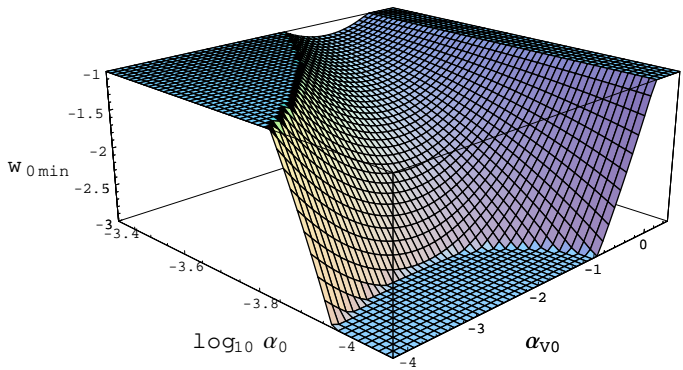,width=9cm}}
 \caption{Minimum value of $w_0$ as a function of $(\alpha _0,\alpha
 _{{\rm v}0})$ for $\sigma _0<10^{-3}$.} \label{wsr}
\end{figure}

We now come back to our master equation~(\ref{wapp}) and analyze the
two regimes where $1+w_0$ can become negative and even large (in
absolute value). The first regime is the natural extension of
quintessence models to scalar-tensor theories~\cite{uqu} and
corresponds to the situation in which the field is decelerating, \ie
$\varphi_{*0}''\ll\varphi_{*0}'$, so that $\alpha_0\varphi_{*0}''$ is
negligible in Eq.~(\ref{wapp}).  Clearly, this requires
$\beta_0>1$. Such a regime, that cannot be reached in a JFBD model,
has the advantage to exhibit an attraction mechanism toward
GR~\cite{damnor} so that $\alpha_0$ can dynamically be made
small. Since, in this situation, $\beta _0$ is not close to $-1$,
Eq.~(\ref{phiprim0}) implies that $\varphi _{*0}'\sim \sigma
_0/[2\alpha _0(1+\beta _0)]$ and Eq.~(\ref{phippapprox}) leads to $\xi
_0\sim \beta _0\sigma _0^2/[2\alpha _0^2(1+\beta _0)]$. On the other
hand, from the Klein-Gordon equation, one obtains that $\varphi
_{*0}'\sim -(\alpha _0X_0+\alpha _{{\rm v}0})/(2+X_0)$ and, in order
to fulfill the condition $\varphi _{*0}'<\sqrt{3}$, one must have
$\vert \alpha _{{\rm v}0}\vert \lesssim4.2$. The two expressions for
$\varphi_{*0}'$ can be used to infer what $\beta _0$ is and, if we
insert the result in Eq.~(\ref{wapp}), one gets
\begin{equation}\label{wsrapp}
 3\Omega _{_\de}^0(1+w_0)\simeq
 \frac{\sigma _0\alpha_{{\rm v}0}} {(2+X_0)\alpha _0}<0\,,
\end{equation}
in the most interesting limit where $\alpha_{{\rm v}0}\gg\alpha_0X_0$
since, otherwise, $\vert 1+w_0\vert \sim\mathcal{O}(\alpha_0^2)$. They
are two ways to interpret Eq.~(\ref{wsrapp}). Either it gives $w_0$ in
terms of $\alpha _0$ and $\alpha _{{\rm v}0}$, assuming that
$\sigma_0$ is known. Or it provides, for fixed $\alpha _0$ and $\alpha
_{{\rm v}0}$, the minimum value that $w_0$ can reach assuming, as is
the case now, that only an upper bound on $\sigma _0$ is
available. Fig.~\ref{wsr} depicts the above-mentioned minimum value as
a function of $(\alpha _0,\alpha _{{\rm v}0})$.

\par

At this point several remarks are in order. Firstly, it is unclear
whether the above situation can be realized without some fine-tuning
in a realistic model, where both the coupling function and the
potential are running away in order to have attraction towards GR and
insensitivity to the initial conditions~\cite{pietroni}.  Let us
illustrate this point on the following particular example.  Consider
the case where $\alpha \propto {\rm e}^{-\lambda \varphi _*}$ so that
$\beta _0=-\alpha _0\lambda$. Since we need $\alpha _0\ll 1$ and, at
the same time, $\beta _0>1 $, this means that $\lambda \gg 1 $ which
may be considered as unnatural. Secondly, one can reverse the logics
and study what a detection of $w_0<-1$ would imply on scalar-tensor
theories (assuming, of course, that a slow-rolling $\varphi _*$ is
causing the acceleration of the expansion).  Besides the fact that the
condition $\beta _0>1$ would drastically improve the current limit on
$\beta _0$ and, in particular, exclude GR, it is also interesting to
remark that this would link $w_0$ to the time variation of the Newton
constant through the expressions $\sigma _0^2\simeq 6\Omega
_{_\de}^0(1+w_0)(1+\beta _0)^2\alpha_0^2/(1-\beta _0)$ and
$\xi_0\sim3\Omega_{_\de}^0
\beta_0(\beta_0+1)(w_0+1)/(1-\beta_0)\lesssim 6\times10^2$, a limit
sharper than the bound set by local experiments. Finally, it is
interesting to notice that $w_0<-1$ can be obtained even if the
potential energy is negligible (in GR this would correspond to
$w_0\sim1$). In this case, one can show that $\varphi_{*0}'\sim
(3\Omega_{_\de}^0)^{1/2}$ and
$\varphi_{*0}''\sim-3\Omega_{\mat}^0\varphi_{*0}'/2$ so that
$\alpha_0\varphi_{*0}''$ is indeed negligible in Eq.~(\ref{wapp}).
Then, the equation of state can easily be obtained and reads
$1+w_0\sim 2(1-\beta _0)$. Again $\beta_0$ is bounded by the
constraint on $\sigma_0$, $1+\beta_0< \sigma_0/(\alpha_0\sqrt{12\Omega
_{_\de}^0})$, so that a large value of $w_0$ requires
$\sigma_0\ll\alpha_0$.

\par

The second regime corresponds to the case where $\alpha\varphi_{*}''$
dominates in Eq.~(\ref{wapp}). As can be seen in
Eq.~(\ref{phippapprox}), this is possible if $\xi_0$ is large and
positive and/or if $\beta _0\sim -1$. Strictly speaking, one should in
fact consider the limit $\beta_0 \rightarrow -(1+\alpha _0^2)$
corresponding, when $V=0$, to the Barker theory~\cite{barker} in which
$A=\cos\varphi_*$ and $G_\cav$ constant ($\sigma=\xi=0$, whatever the
value of $\varphi_*'$ and $\varphi_*''$). Since
$\alpha=-\tan\varphi_*$, the cosmological evolution drives the theory
away from GR unless a potential keeps $\varphi_*$ close to 0 until the
last $e$-folds inducing a large variation of
$\alpha\varphi_*''\sim\mathcal{O}(1)$ in order to have $1+w_0<0$. Such
a model seems very contrived unless the potential exhibits a slope
discontinuity recently.

\par

Let us now assume that $\xi_0\sim 0$ since, from dimensional analysis,
one expects $\xi _0\sim \sigma _0^2 \lesssim 2\times 10^{-3}$. In the
limit $\beta _0\rightarrow -1-\alpha _0^2$, Eq.~(\ref{phippapprox})
leads to
$\alpha\varphi_{*0}''\sim2\alpha_0^2\beta_0^2\varphi_{*0}^{\prime2}/(1+\beta
_0+\alpha _0^2)$ and $w_0$ diverges when $\beta _0\rightarrow
-1-\alpha _0^2$. However, the constraint $\varphi_*'<\sqrt{3}$ implies
that $1+\beta _0+\alpha _0^2 \gta \sigma_0/(2\sqrt{3}\alpha_0)$ and,
hence, the smallest value that can be obtained is $(1+w_0)_{\rm
min}\sim -8\sqrt{3}\alpha _0^3/(\Omega_{_\de}^0\sigma_0)$. In
particular, $w_0\ll -1$ is perfectly possible in this regime if
$\sigma _0\ll \alpha _0^3$.  Finally, let us see what this regime
implies in terms of model building. The Klein-Gordon equation implies
that $2\varphi_{*0}''/3\sim-\alpha_{{\rm v}0}/(2+X_0)$ so that the
slope of the potential must be very large, $\alpha_{{\rm
v}0}\gg\alpha_0^{-1}$. Moreover since, in this regime, $\beta _0<0$
the theory is driven away from GR and it follows that the potential
must be tuned in order to prevent this drift.


As a conclusion, let us summarize our main findings. Firstly, we have
confirmed that $1+w_0$ is not positive definite in scalar-tensor
theories even if all the matter energy conditions are
satisfied. Secondly, we have established under which conditions
$1+w_0$ can become negative given the local constraints coming from
solar system and pulsars measurements (in the case of chameleon
models~\cite{cameleon}, it has been argued that $\alpha_0$ can be of
order unity today, a case where large negative values of $1+w_0$ can
be achieved more easily). We have shown that getting a non negligible
deviation from $-1$ necessarily implies a non vanishing $\beta _0$
(except if $\xi _0$ is large), a situation that cannot be reached in
the JFBD case. Thirdly, in terms of model building, we have
demonstrated that this corresponds either to $\beta _0>1$, a situation
where the scalar field is slow-rolling today and the coupling constant
is very steep or to $\beta _0\sim -1$ (or $\xi _0$ large) where the
slope of the potential is very large. Fourthly, we have also shown how
a measurement of $w_0<-1$ could improve the local constraints on the
deviations from GR. This highlights the complementarity~\cite{compl}
between cosmological and local tests of gravity.
Finally, let us remark that we have only taken into
account the local constraints. A next step would be to reconstruct the
redshift evolution of the models and to show that they are not
pathological~\cite{gefpol}. {\bf Acknowledgments:} we thank
G.~Esposito-Far\`ese for many enlightening comments and discussions.


\end{document}